\documentclass[conf]{new-aiaa}
\usepackage[utf8]{inputenc}
\usepackage{textcomp}

\usepackage{graphicx}
\usepackage{xcolor}
\usepackage{color}
\usepackage{amsmath}
\usepackage[version=4]{mhchem}
\usepackage{siunitx}
\usepackage{longtable,tabularx}
\title{Thermal Explosion Characteristics of a Gelled Hypergolic Droplet}

\author{Prabakaran Rajamanickam \footnote{PhD candidate, Department of Mechanical and Aerospace Engineering.}}
\affil{University of California San Diego, La Jolla, CA, 92093}

\newcommand{\La}{\mathcal{L}}
\newcommand{\dd}{\mathrm{d}}
\newcommand{\erfc}{\mathrm{erfc}}

\newcommand{\pfr}[2]{\ensuremath{\frac{\partial #1}{\partial #2}}}

\newcommand{\ep}{\epsilon}
\newcommand{\blue}[1]{\textcolor{black}{#1}}

\begin{document}

\maketitle

\begin{abstract}
When a sphere of one reactant is placed in the medium of another reactant with which it is hypergolic, a chemical reaction (modeled here as a zeroth-order one-step irreversible Arrhenius reaction) occurs at the common interface of the two reactants, and the heat generated at the interface then is transmitted away from it by thermal conduction. Depending on the nature of the problem, the system may approach an explosive mode, or it may settle into a steady-state mode. The critical condition defining the transition between these two states is determined analytically. In particular, \blue{explicit formulas for the ignition delay time for the explosive mode are provided for two limits, one in which an appropriately defined Damk\"ohler number is large and other in which it is closer to the critical conditions}. This is accomplished here by deriving and solving an integral equation for the time evolution of the interface temperature, employing activation-energy asymptotics.
\end{abstract}

\section*{Nomenclature}

{\renewcommand\arraystretch{1.0}
\noindent\begin{longtable*}{@{}l @{\quad=\quad} l@{}}
$a$  & fuel radius \\
$T_{Fo}'$ &  initial fuel temperature \\
$T_{Oo}'$& initial oxidizer temperature \\
$T_{Io}'$ & initial interface temperature \\
$\lambda_F$ & thermal conductivity of fuel \\
$\lambda_O$ & thermal conductivity of oxidizer \\
$r_F$ & thermal responsivity of fuel \\
$r_O$ & thermal responsivity of oxidizer\\
$q_s$ & surface heat generation rate \\
$E$ & activation energy \\
$\bar R$   & universal gas constant \\
$t_{\mathrm{ign}}$ & ignition delay time \\
$\delta$ & Damk\"ohler number \\
$\gamma$ & non-dimensional difference of initial temperatures \\
$R=r_O/r_F$ & oxidizer-to-fuel thermal responsivity ratio \\
$L=\lambda_O/\lambda_F$ & oxidizer-to-fuel thermal conductivity ratio
\end{longtable*}}

\section{Introduction}

\lettrine{T}{he} \blue{definitive property of hypergolic reactants is that they do not require a hot ignition source to initiate the chemical reaction. Despite its hazardous nature, there is renewed interest in hypergolic propellants, as may been from recent technical reports~\cite{williamsreport,heister2014} and the results of many studies referenced therein, in part, due to the fact it is easier to handle these reactants once they have undergone gelation with additives. In addition, the gellant additives can also be used to vary the overall transport properties of the medium. In actual rocket propulsion, these reactants are injected as jets into the combustion chamber, where reactants mainly interact with each other through droplet-droplet collisions. From the viewpoint of theoretical modeling, however, in order to clarify the hypergolicity, a situation in which one gelled reactant is syringed or, dropped into a pool of another gelled reactant may be considered first, as in laboratory experiments.}

\blue{In this connection},  a particular model for ignition of hypergolic gelled propellants was presented in which two hypergolic reactants of infinite extent in contact with each other at a common planar interface was studied by Williams~\cite{Williams} in a relatively recent paper. It was found there that explosion occurs always at a finite time. Further attempts to improve this model~\cite{hassid2013,castaneda2018} are still found to be confined to the Cartesian geometry considered in~\cite{Williams}. If, however, one or both of the reactants are of finite extent, then it may be expected that the event of ignition itself may not occur in the first place owing to the fact that under certain conditions a steady state can be reached for the heat-transport process. The transition between thermal runaway and steady state, in typical combustion processes, is usually discrete, a well-known phenomenon associated with the large activation energy of the reaction. Hence, a parameter i.e., a Damk\"ohler number, characterizing this discrete transition, has to be introduced to consider reactants of finite extent. 

In this paper the problem in which one of the reactant is of finite extent is considered, with other simplifications and assumptions being essentially the same as in the earlier work~\cite{Williams}. For simplicity, the finite-sized reactant is assumed to be spherically symmetric, although predictions are expected to be valid for other similar geometries with characteristic sizes the same as that of the radius of the sphere.

A volterra-type nonlinear integral equation can be derived for problems of these kind, as can be seen from classical ignition theories of solid propellants~\cite{williams1966,williams1969,williams1970} and from problems of heat (mass) transfer with surface radiation (reaction)~\cite{mann,chambre,acrivos1957}. Here, a similar integral equation governing the time evolution of the interface temperature will be derived for spherically symmetric heterogeneous systems. Given the number of approximations introduced in this elementary study directed towards the understanding of influences due to finite reactant-proportions, following~\cite{Williams}, a complete solution to this integral equation is not attempted here, but an approximate solution, bringing out the essential features of this equation, for two extreme cases of Damk\"ohler number, is presented. The required critical values for the Damk\"ohler number are also derived from this integral equation directly.

\section{The governing equations and the boundary conditions of the problem} \label{sec:formulation}
The problem can be described as follows. At time $t=0$, a drop of fuel of radius $a$ and temperature $T_{Fo}'$ is placed in an infinitely large oxidizer environment, maintained at a temperature $T_{Oo}'$. Then with the interface reaction assumed to obey a zeroth-order one-step irreversible Arrhenius reaction, it is required to find the temperature at all locations for all further instants $t>0$; in particular, the time history of the interface temperature is of central interest.

In view of the simplifications assumed, which are the same as those of~\cite{Williams}, the governing equation for the temperature $T_i'$ of species $i$ is simply the heat equation,  
\begin{equation}
    \pfr{T_i'}{t} = \frac{\alpha_i}{r^{2}}\pfr{}{r}\left(r^{2}\pfr{T_i'}{r}\right), \qquad i = F,O
\end{equation}
where $\alpha_i$ is the thermal diffusivity of species $i$ and $r$ is the distance measured from the center of the fuel-drop. The initial and boundary conditions are given accordingly by
\begin{align}
    t=0: &\qquad T_F' = T_{Fo}',\, (0<r<a),\qquad T_O' =T_{Oo}',\, (a<r<\infty),\\
    t>0: &\qquad \pfr{T_F'}{r}=0,\, (r=0), \qquad T_O' =T_{Oo}',\, (r\rightarrow\infty).
\end{align}
Since the heterogeneous reaction occurs at the interface $r=a$, the conditions that need to be satisfied at the interface are
\begin{equation}
     T_F'=T_O'=T_I', \quad \text{and}\quad\lambda_F \pfr{T_F'}{r} - \lambda_O \pfr{T_O'}{r}  =  q_s e^{-E/\bar RT_I'}, \label{sec2eqn4}
\end{equation}
where $T_I'(t)$ is the interface temperature, $\lambda_i$ is the thermal conductivity of species $i$, and $q_s$ is the rate of heat release per unit surface area. Here, $E$ is the activation energy of the one-step reaction and $\bar R$ is the universal gas constant. The second interface condition is obtained by integrating the time-dependent energy equation across the interface, taking the limit of vanishing interface thickness with the assumption that $q_s$ is finite. \blue{In the above  model, fuel and oxidizer are distinguished purely by geometrical effects, whereas the one-step model chemistry is completely symmetric with respect to fuel and oxidizer. If the geometrical role of fuel and oxidizer are interchanged, all corresponding results and discussions should also be interchanged.}

The linear problem, free of interface heat-release, obtained by setting $q_s=0$ in equation~\eqref{sec2eqn4}, has been solved by Brown~\cite{Brown} for constant initial temperatures and by Konopliv and Sparrow~\cite{Sparrow} for arbitrary initial temperature distributions.

As a matter of fact, the spatial temperature profile evolves on two different time scales, namely, the diffusion time scale that is important in most of the region and the chemical time scale that is assumed to exert an effect only at the interface. These two time scales are different from each other unless their ratio (defining a Damk\"ohler number) is of order unity. Numerical methods that try to solve the above set of equations must be such that the fastest of the two time scales be well resolved. But, for problems of this nature, the complexity associated with the two time scales can be circumvented by separating the interface problem from the outer diffusional problem in a sense that will be clear as we proceed further.
 
To facilitate the outer diffusion problem first, we introduce the dimensionless variables and parameters
\begin{equation}
    \tau = \frac{\alpha_F t}{a^2}, \quad y=\frac{r}{a}, \quad T_i = \frac{T_i'}{T_{Fo}'},\quad \alpha=\frac{T_{Oo}'}{T_{Fo}'}, \quad L = \frac{\lambda_O}{\lambda_F}, \quad R=\frac{r_O}{r_F}, \quad Q = \frac{aq_s}{\lambda_F T_{Fo}'}, \label{nondim1}
\end{equation}
where $r_i=\lambda_i/\sqrt\alpha_i$ is the thermal responsivity (or thermal effusivity) of species $i$. The governing equations consequently become
\begin{equation}
     \pfr{T_F}{\tau} = \frac{1}{y^{2}}\pfr{}{y}\left(y^{2}\pfr{T_F}{y}\right), \quad  \pfr{T_O}{\tau} = \frac{L^2}{R^2y^{2}}\pfr{}{y}\left(y^{2}\pfr{T_O}{y}\right) \label{nond1}
\end{equation}
while the corresponding initial and boundary conditions change to
\begin{align}
    \tau=0: &\qquad T_F = 1,\, (0<y<1),\qquad T_O =\alpha,\, (1<y<\infty),\\
    \tau>0: &\qquad \pfr{T_F}{y}=0,\, (y=0), \qquad T_O =\alpha,\, (y\rightarrow\infty).
\end{align}
Finally, the interface conditions for $\tau>0$ can be written as
\begin{equation}
    y=1: \qquad  T_F=T_O=T_I, \qquad \pfr{T_F}{y} -L \pfr{T_O}{y}  =   Qe^{-E/\bar RT_{Fo}'T_I}. \label{sec3eqn4}
\end{equation}
The above problem possess a non-trivial steady solution unique to the spherical geometry considered here, while no such steady solutions can be found for configurations with cylindrical or planar symmetry. The discussion of such solutions will be postponed to section V, since it seems helpful to deduce results in terms of activation-energy asymptotics first.

\section{A general method of solution and the derivation of an integral equation governing the interface temperature} \label{sec:solution}
Since the nonlinear term enters only through a boundary condition, methods for linear problems can be applied here to the extent that they do not involve the interface condition exhibiting the nonlinear behavior. For the same reason, there are several ways to arrive at the same result, of which the technique of Laplace transforms is exploited in this paper, as was done in references~\cite{Williams,williams1966,williams1969,williams1970,mann,chambre,Brown,Sparrow}. The Laplace transforms of the dependent variables are defined as
\begin{equation}
    \tilde T_i = \mathcal{L}(T_i)=\int_0^\infty T_i\, e^{-s\tau}\, \dd\tau, \qquad i = F,O. \label{lapT}
\end{equation}

The governing equations then reduce to ordinary differential equations,
\begin{align}
    \frac{\dd^2\tilde T_F}{\dd y^2} + \frac{2}{y}\frac{\dd\tilde T_F}{\dd y} =  s\tilde T_F -1, \quad \frac{\dd^2\tilde T_O}{\dd y^2} + \frac{2}{y}\frac{\dd\tilde T_O}{\dd y} = \frac{ R^2}{L^2}(s\tilde T_O-\alpha),
\end{align}
whose solutions satisfying the conditions away from the interface are
\begin{align}
    \tilde T_F = \tilde F_F(s) \frac{\sinh \sqrt{ s}y}{y} + \frac{1}{s} ,\qquad
    \tilde T_O = \tilde F_O(s) \frac{e^{-y\sqrt{s}R/L}}{y} + \frac{\alpha}{s} \label{Tlap}
\end{align}
where $\tilde F_F(s)$ and $\tilde F_O(s)$ are constants of integration. Continuity of temperature at $y=1$ implies
\begin{equation}
  \tilde F_F\sinh \sqrt{ s} + \frac{1-\alpha}{s}= \tilde F_O e^{-\sqrt{ s}R/L}, \label{continuity}
\end{equation}
whereas the flux-jump~\eqref{sec3eqn4} at the interface leads to
  \begin{align}
     \tilde F_F\sinh\sqrt{ s}(\sqrt{ s}\coth\sqrt{s} -1) + \tilde F_O e^{-\sqrt{ s} R/L}(\sqrt{ s} R + L) = Q\int_0^\infty e^{-E/\bar RT_{Fo}'T_I} e^{-s\tau}\, \dd\tau. \label{fluxjump}
 \end{align}
By eliminating $\tilde F_O $ between the foregoing two equations we obtain an equation for $\tilde F_F$, 
\begin{equation}
    \blue{\sqrt{s} \tilde F_F \sinh \sqrt{s} \left(\coth\sqrt s + R + \frac{L-1}{\sqrt s}\right)+ (1-\alpha)\left(\frac{R}{s} + \frac{L}{\sqrt s}\right) = Q\int_0^\infty e^{-E/\bar RT_{Fo}'T_I} e^{-s\tau}\, \dd\tau. } \label{fluxjumpff} 
\end{equation}
This completes the solution for the outer diffusion problem  since $T_F$ and $T_O$ in principle can be expressed as quadratures by inverting~\eqref{lapT}, with $\tilde F_F$ determined from an integral equation that governs the inner interface temperature.

\subsection{\blue{Determination of initial interface temperature}}
The interface temperature, although discontinuous at $t=0$, will become continuous instantaneously for $t>0$ as a necessary consequence of the heat equation, such that we can speak of a unique interface temperature as $t\rightarrow 0^+$. Also, since the heat release by chemical reaction has not begun at these initial instants, the initial interface temperature can be obtained from equation~\eqref{fluxjumpff} after removing its right-hand side, \blue{to result in}
\begin{equation}
    \blue{s\tilde F_F \sinh\sqrt s = - \frac{(1-\alpha)(R+L/\sqrt s)}{\coth\sqrt s + R + (L-1)/\sqrt s}.}
\end{equation}
\blue{By definition, $\tilde F_F \sinh\sqrt s$ is the Laplace transform of $T_I-1$ and to find $T_{Io}-1$, where $T_{Io}$ is the initial interface temperature in non-dimensional form, we take the limit $s\rightarrow \infty$ and apply initial-value theorem,} yielding
\begin{equation}
    T_{Io} = \frac{1+\alpha R}{R+1}.
\end{equation}
The dimensional initial interface temperature $T_{Io}'$ is simply a weighted average of initial temperatures with thermal responsivities being the weight factors. This result, the same as that of the planar problem~\cite{Williams}, is in fact valid for arbitrary interfaces that are smooth.

\subsection{\blue{The integral equation}}

Following Williams~\cite{Williams}, next we study the evolution of the departure of the interface temperature from its initial value, i.e., we define $F=T_I-T_{Io}$, so its Laplace transform $\tilde F = \La(F)$ becomes $\tilde F = \tilde F_F\sinh \sqrt s + R(1-\alpha)/[s(R + 1)]$. Equation~\eqref{fluxjumpff} can be re-written in terms of $\tilde F$ as
 \begin{align}
        \sqrt s \tilde F \left(\coth\sqrt s + R  \right)  + \frac{R(1-\alpha)(1-\coth\sqrt s)}{\sqrt s(R+1)}  + \frac{(1-\alpha)(L+R)}{s(R+1)} + (L-1)\tilde F
        = Q \int_0^\infty e^{-E/\bar RT_{Fo}'(F + T_{Io})}e^{-s\tau}\, \dd\tau. \label{lap1}
\end{align}
 Taking the inverse Laplace transform using Bateman's tables~\cite{Bateman1954} gives
 \begin{align}
     \int_0^\tau \frac{\dd F(\tau')}{\dd\tau'} \left[ \vartheta_3\left(0|i\pi(\tau-\tau')\right) + \frac{R}{\sqrt{\pi(\tau-\tau')}}\right]\dd\tau' + \frac{R(1-\alpha)}{R+1}\left[\frac{1}{\sqrt{\pi \tau}} -\vartheta_3(0|i\pi\tau) + \frac{L}{R} + 1\right]
    + (L-1)F \nonumber \\
    = Qe^{-E/\bar RT_{Fo}'[F+T_{Io}]}, \label{intnorm}
 \end{align}
  where $\vartheta_3(0|i\pi\tau)$ is one of the Jacobi theta functions, in which the argument $i\pi\tau$ often is called the lattice parameter. The preceding equation is the required integral equation for $F(\tau)$ subjected to the condition $F(0)=0$. 

\section{The application of activation-energy asymptotics} \label{sec:AEA}

As previously discussed, evolution of the interface temperature should be scaled with the chemical time. In large-activation-energy asymptotics, the small parameter of expansion can be defined as $\ep=\bar R T_{Io}'/E\ll 1$. From the dimensional equations (1)-(4), it is clear that the characteristic heat-conduction time and the characteristic explosion time, respectively, are given by
\begin{equation}
    t_c = \frac{a^2}{\alpha_F} \quad \text{and} \quad t_e = \frac{a\epsilon T_{Io}' \lambda_F}{ q_s e^{-1/\epsilon} \alpha_F}.
\end{equation}
The non-dimensional parameters and variables required for activation-energy asymptotics become
 \begin{equation}
     \delta = \frac{t_c}{t_e}, \quad  \varpi = \frac{\pi t}{t_e} = \delta \pi\tau , \quad \psi = \frac{F}{\epsilon T_{Io}} = \frac{F(R+1)}{\epsilon(1+\alpha R)}, \quad \gamma = \frac{1-\alpha}{\ep(1+\alpha R)}, \label{nondim2}
 \end{equation}
 where $\delta$ is the Damk\"ohler number. The parameter $\gamma$, measuring the relative difference between the initial temperatures of the two reactants, can range from negative values of the orders of tens to positive values of the same orders, for the typical values $\alpha\sim O(1)$ and $\ep = 0.01-0.1$.  

 With the scales introduced in this section, equation~\eqref{intnorm} becomes
  \begin{align}
     \int_0^\varpi\frac{\dd\psi(\varpi')}{\dd\varpi'}\left[ \vartheta_3\left(0\Big |\frac{i(\varpi-\varpi')}{\delta}\right) + \frac{R\sqrt\delta}{\sqrt{\varpi-\varpi'}}\right] \dd\varpi' + R\gamma \left[\frac{\sqrt\delta}{\sqrt\varpi} -\vartheta_3\left( 0 \Big |\frac{i\varpi}{\delta}\right) + \frac{L}{R} + 1\right]
     +  (L-1) \psi \nonumber \\
     = \delta  e^{\psi/(1+\epsilon \psi)} \approx \delta e^\psi, \label{intfinal}
 \end{align}
where, on the premise of activation-energy asymptotics, the Arrhenius term at the end has been linearized following Frank-Kamenetskii~\cite{Frank}. The explicit dependence on $\ep$ has thus been removed, thereby leaving  $R$, $L$, $\gamma$, and $\delta$ as the four independent parameters of the problem. 

The particular theta function that appears in the integral equation can be represented by two different forms\footnote[2]{The two representations are related by the identity $\vartheta_3(0|ix^{-1})=\sqrt{x}\vartheta_3(0|ix)$.},
  \begin{equation}
     \vartheta_3\left( 0 \Big |\frac{i\varpi}{\delta}\right) = 1 + 2 \sum_{n=1}^\infty e^{-\pi n^2\varpi/\delta}=\frac{\sqrt{\delta}}{\sqrt{\varpi}}\left[1+ 2 \sum_{n=1}^\infty e^{-\pi n^2\delta/\varpi}\right].
 \end{equation}
 where the second representation is obtained  by taking the Jacobi imaginary transformation (cf.~\cite{Whittaker}, p.~475) of the first representation. These two representations will be useful in deriving the limiting forms of the theta function in various limits.

\section{The steady solution and the Frank-Kamenetskii parameter} \label{sec:steady}
For given values of $R$, $L$ and $\gamma$, a steady solution is expected to occur only for $\delta\leq\delta_c$, where $\delta_c$ is the critical Damk\"ohler number, also known as the Frank-Kamenetskii parameter~\cite{Frank}.
Taking the limit $\varpi\rightarrow\infty$ in equation~\eqref{intfinal}, keeping other parameters fixed, the steady solution $\psi_s$ is found to satisfy the following equation\footnote[3]{The temperature distribution in steady state, outside the interface is given by $T_F=T_I$ and $T_O = (T_I-\alpha)/y + \alpha$, where $T_I = T_{Io}(1+\ep\psi_s)$.}
\begin{equation}
    \psi_s + \gamma = \frac{\delta}{L} e^{\psi_s}. \label{steady1}
\end{equation}
 Except for the appearance of $\gamma$, the above equation is similar to one obtained by Semenov~\cite{semenov1928} in the thermal-explosion theory of well-mixed reactants. One may also notice that, in steady state, the heat liberated by the interface reaction is lost only to the oxidizer as in Semenov's case, where heat loss occurs only at the enclosing wall, but for a different reason, i.e. due to the assumption of large conductivity for the reactant mixture in the Semenov analysis.

An explicit solution can be written for $\psi_s$ in terms of Lambert W function~\cite{Lambert},
\begin{equation}
   - \psi_s = \gamma  + W\left(-\frac{\delta e^{-\gamma}}{L}\right).\label{steady2}
\end{equation}
  Since the Lambert W function is real only when its argument $x\geq -1/e$, we immediately determine the condition for existence of a real, steady interface temperature, written in terms of the critical Damk\"ohler number, to be
\begin{equation}
    \delta \leq \delta_c = L e^{\gamma-1}. \label{critical}
\end{equation}
 For cases in which both fuel and oxidizer have the same initial temperature, the Frank-Kamenetskii parameter $\delta_c=L/e$ is an order-unity number. By way of contrast, if the fuel sphere has an initial temperature which is much hotter than the oxidizer ($\alpha \ll 1$), then $\delta_c\sim e^{1/\ep}$, i.e. in the limit of vanishing $\ep$, the system never goes to explosive mode, while, in the opposite case ($\alpha \gg 1$), the system never reaches a steady state, since in this limit $\delta_c\sim e^{-1/\ep}$. 

 \begin{figure}[ht]
\centering
\includegraphics[scale=0.6]{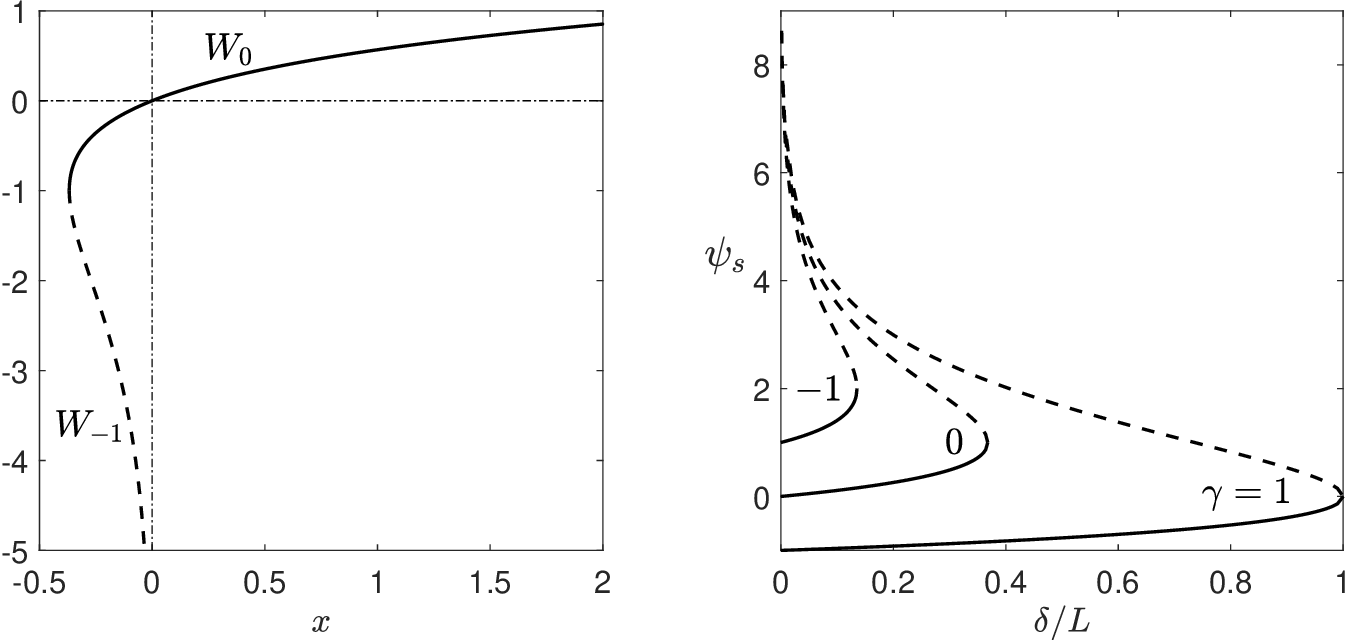}
\caption{The left-hand side plot shows the real-valued figure of Lambert W-Function, comprising of two branches, the principal branch (solid curve) and the lower branch (dashed curve), whereas the right-hand side plot draws the curve for steady solution, evaluated using both branches of the Lambert function.}
\label{fig:steady}
\end{figure}

For obvious reasons, the argument of the Lambert W function in~\eqref{steady2} is nonpositive in all possible situations, while the function $W(x)$ itself has two branches in that range $-1/e\leq x<0$, the upper (principal) branch, $W_0(x)$ and the lower branch, $W_{-1}(x)$ (see left plot of Fig.~\ref{fig:steady}), of which one has to be selected based on physical grounds. It seems that the principal branch obeys the fact that a steady interface temperature, whenever it exists, should increase for increasing values of $\delta$ when other parameters are kept fixed, a typical expectation which follows directly from S-shaped ignition-extinction curve. This can be seen to be true from the right plot of Fig.~\ref{fig:steady}, where steady solutions are plotted as functions of $\delta/L$ for different possible values of the Frank-Kamenetskii parameter via $\delta_c/L$ for both branches of $W(x)$. By comparing this figure with the typical curves of thermal-explosion theory, we can identify $W_0$ to be the stable branch and $W_{-1}$ as unstable.

\section{A simplified integral equation applicable in the range $\delta \gg 1$ and its approximate solution} \label{sec:integral}
 
 Now that we have a well-defined criterion for the existence of steady solutions, we can turn our attention to unsteady problems and consider the limit of large Damk\"ohler numbers first. As $\delta\rightarrow \infty$, a thermal runaway is expected to occur in a short interval of time from the inception, with solutions approaching the planar problem considered by Williams~\cite{Williams}\footnote[4]{A more precise statement would be to define $\delta-\delta_c\rightarrow \infty$ as the planar limit, rather than $\delta\rightarrow \infty$. For example, eqn.~\eqref{volterra1} does not always produce an explosive solution; $\delta$ can be less than $\delta_c$.}. For the planar geometry, ignition occurs at times of order $\varpi\sim\delta^{-1}$. Thus, the ratio $\varpi/\delta$ can at most be of the order $\delta^{-2}$ and hence the theta function can be approximated as $\vartheta_3(0|i\varpi/\delta) \approx \sqrt{\delta/\varpi}$ with exponentially small corrections.
 
 By defining a new time coordinate $\eta =\delta\varpi/(R+1)^2$ and by introducing two new parameters $\Lambda=\gamma(R+L)/\delta$ and $k=(L-1)/\delta$, the integral equation~\eqref{intfinal}, in the limit of large Damk\"ohler number, can be shown to reduce to
 \begin{equation}
     \int_0^\eta \frac{\dd\psi}{\dd\eta'}\frac{\dd\eta'}{\sqrt{\eta-\eta'}} + \Lambda + k\psi = e^\psi, \label{volterra1} 
 \end{equation}
 where the function $\psi$ inside the integral should be interpreted as $\psi=\psi(\eta')$. Due to the exponentially small corrections, the above equation will not differ markedly from~\eqref{intfinal} even for moderately large $\delta$; for example, the error is already of order $10^{-4}$ when $\delta=3$. For this reason, although strictly speaking, in the asymptotic sense, $\Lambda$ and $k$ should be treated as first-order corrections, we may as well treat them as order unity quantities since this may be true for large but finite values of $\delta$.

 By the above arguments, we can see that the \blue{distinguished} double limit $(\Lambda,k)\rightarrow (0,0)$ is planar in character, and in this limit, the simplified integral equation reduces to the Williams equation~\cite{Williams}. The conditions under which $\Lambda$ and $k$ vanishes can be inferred directly from their definitions.

The simplified integral equation is not soluble in terms of elementary functions, whence it will be solved locally both for small times and near the point where the solution diverges indefinitely. An approximation to a uniform solution will be constructed out of these two solutions by the method of patching. Only those values of $\Lambda$ and $k$ for which these local solutions are reasonably accurate and for which $\delta>\delta_c$, will be considered.

For $\eta\ll 1$, a power-series expansion leads to
\begin{equation}
    \psi(\eta) = \frac{2p}{\pi}\sqrt{\eta} + \frac{ps}{\pi}\eta + \frac{4p(2p+\pi s^2)}{3\pi^3}\eta^{3/2} + \frac{p[2p^2 + ps(3\pi+4) + 2\pi s^3]}{4\pi^3} \eta^2+\cdots \label{initial1}
\end{equation}
 where $p=1-\Lambda$ and $s=1-k$ have been introduced for shortness. 

 As far as the terminal profile is concerned, obtaining a neat perturbation series like the one for the initial profile is non-trivial; therefore an approximate solution will be derived in the following. Let $\eta=c$ be the point at which the solution diverges. Then in the neighborhood of $\eta=c$, the upper limit $\eta$ appearing in the integral in Eqn.~\eqref{volterra1} can be replaced by $c$ and the lower limit of integration itself can be shifted to $\eta$ since the major contribution to the integral term comes from this neighborhood. A simple consideration shows then that the integral equation~\eqref{volterra1} can be approximated by the  differential equation $ \pi\sqrt{c-\eta}\dd\psi/\dd\eta + \Lambda  + k\psi = e^\psi$.\footnote[5]{The factor $\pi$ is easy to obtain if one recognises that one can always find a region near $\eta=c$ where $\Lambda$ and $k\psi$ are negligible when compared to the other two terms. With $\Lambda$ and $k$ neglected then, only $\pi$ is found to be consistent for the solution of the differential equation and the leading-order solution of the integral equation~\eqref{volterra1} to be the same.} The equation has to integrated with the condition that $\psi$ becomes infinite at $\eta=c$. A straightforward integration shows then that
 \begin{equation}
     \frac{2}{\pi}\sqrt{c-\eta} = G(\psi;\Lambda,k) \equiv \int_\psi^\infty \frac{\dd\psi'}{e^{\psi'} -\Lambda - k\psi'}. \label{terminal}
 \end{equation}
 The function $G$, in general, has to be calculated numerically except for the case $k=0$, for which the above equation simplifies to $\Lambda e^{-\psi}=1-e^{-2\Lambda\sqrt{c-\eta}/\pi}$, which as it must, reduces to eqn.(23) of~\cite{Williams} for small values of $\Lambda$. Although $G$ is well-behaved since it tends to $e^{-\psi}$ as $\psi\rightarrow \infty$, it may become singular in a logarithmic fashion depending on the values of $\Lambda$ and $k$ if $\psi$ is not sufficiently large. For the values considered for $(\Lambda,k)$ in this paper and for the domain where both initial and terminal solution are accurate enough, $G$ is non-singular.

The problem remains to determine the constant $c$ by matching the initial and the terminal solution. In order to do that, $\eta$ will be treated as a function of $\psi$. An intermediate point $\psi=\psi_p$ is assumed to exist where both the initial profile~\eqref{initial1} (after solving for $\eta=H(\psi;\Lambda,k)$) and the terminal profile~\eqref{terminal} have the same value, at which point, the slope of two curves is also assumed to be continuous. This leads to two conditions for the two unknowns, $c$ and $\psi_p$. The patching point $\psi_p$ can be found from an appropriate root of $4\dd H/\dd\psi + \pi^2 \dd G^2/\dd \psi = 0$ and the value of $c(\Lambda,k)$ can be found from $c = H(\psi_p) + \pi^2 G^2(\psi_p)/4$.

 \begin{figure}[ht]
\centering
\includegraphics[scale=0.6]{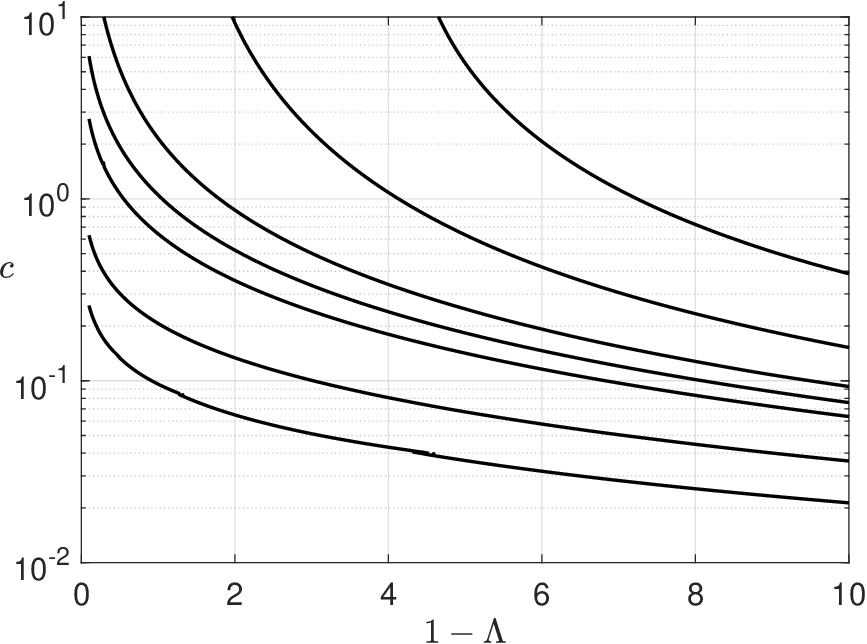}
\caption{Plots of explosion point $c(\Lambda)$ for values of $k=-10,\,-5,\,-1,\,0,\,1,\,3,\,5$ corresponding to bottom curve to top curve, respectively .}
\label{fig:cetap}
\end{figure}

To plot results, we keep first the four terms of~\eqref{initial1} in evaluating $c$ since retaining more terms seems to have little influence on the value of $c$. The curves of $c(\Lambda)$ for fixed values of $k$ are plotted as solid curves in Fig.~\ref{fig:cetap}. The result is increasingly accurate for decreasing values of $c$. Increasing either $\Lambda$ or $k$ evidently delays ignition so that the delay time can become larger or smaller than that of the planar problem depending on values of these two parameters. For Williams~\cite{Williams}, who had kept only the first term in~\eqref{initial1}, the value of $c$ is universal and equal to $c(0,0)$. This value which he found to be $1.5$ seems to decrease and approach unity as we include more terms to the initial solution. \blue{Numerical integration of~\eqref{volterra1} for $(\Lambda,k)=(0,0)$ using the scheme described in~\cite{Williams} showed that the actual value of $c$ is around $0.8$, $20$ percent lower than the predicted value, thereby providing an estimate of the expected magnitude of the error.}

\section{The solution characteristics for sub-critical and near-critical limits}

For simplicity, only the case $\gamma=0$ (equal initial temperatures) is considered in this section. Then from~\eqref{critical}, the Frank-Kamenetskii parameter simply becomes $\delta_c=L/e$, while the steady-state value of $\psi$ at this critical condition is $\psi_s=1$ according to~\eqref{steady2}. Also, the steady-state value of $\psi$ for $\delta<\delta_c$ is $\psi_s= -W_0(-\delta/L)$ from~\eqref{steady2}.

Prior to investigating the near-critical limit $|\delta-\delta_c|\ll 1$, it is helpful to first understand the solution behavior throughout the sub-critical range $\delta<\delta_c$. In the sub-critical regime, the transient solution evolves on time scales of the order of $t_c$, eventually reaching a steady value, and therefore it is appropriate to work with $\tau$ in place of $\varpi$ as the relevant non-dimensional time variable. Here, we are interested only in the asymptotic behavior of the transient solution for large $\tau$. In terms of $\tau$, the integral equation~\eqref{intfinal}, as $\tau\rightarrow\infty$, after introducing $\psi(\tau)=\psi_s + \theta(\tau)$, with $\theta\ll 1$, reduces to
\begin{equation}
    \int_0^\tau \frac{\dd\theta}{\dd\tau'}\frac{R\dd\tau'}{\sqrt{\pi(\tau-\tau')}}  + \mu L\theta =  L(1-\mu)\left(\frac{\theta^2}{2!} + \frac{\theta^3}{3!} + \cdots\right), \label{closesteady}
\end{equation}
where $\mu=1-\psi_s<1$ measures the difference between the steady-state temperature for the critical condition $\delta=\delta_c$ and that for the sub-critical condition $\delta<\delta_c$; $\mu$ vanishes when $\delta=\delta_c$. The linear integral equation that is obtained by neglecting $\theta^2$ and higher-order terms can be solved by Laplace transforms~\cite{Bateman1954}, to give
\begin{equation}
    \theta = \theta_o e^{L^2\mu^2 \tau/R^2} \erfc\left(\frac{L\mu}{R}\sqrt{\tau}\right) \approx \frac{R\theta_o}{L\mu\sqrt{\pi\tau}}, \label{closesteady1}
\end{equation}
where $\theta_o$ is a constant that can be determined numerically. This indicates an algebraic decay to the steady solution which is qualitatively different from the exponential decay (cf.~\cite{kassoy1975}, eqn.14 and~\cite{kassoy1978}, eqn.3.1) observed for Semenov's problem. 

The above asymptotic solution is valid for order-unity values of $\mu$, but in the near-critical limit $\mu\rightarrow 0$, it develops a sequence of non-uniformities as $\tau$ runs to infinity; for instance, the $\theta^2$ term in~\eqref{closesteady} is no longer negligible when $\theta\sim\mu$, that being the order of magnitude when $\tau\sim \mu^{-4}$ according to~\eqref{closesteady1}. A familiar example that is analogous to this situation is the difficulty in obtaining a uniform formula through activation-energy asymptotics for the laminar burning velocity of planar premixed flames, that include both near-stoichiometric and far-from-stoichiometric equivalence ratios. The origin of trouble lies in the expansion about $\psi_s$ for $\psi$, while the correct expansion should have been about unity for near-critical limits.  

The failure of the preceding solution for the near-critical limit forces us to reconsider the inherent balance in~\eqref{closesteady} in order to identify the correct leading-order terms. In trying to reformulate the problem for the near-critical case $\delta_c-\delta\ll 1$, we can also incorporate the super-critical case $\delta-\delta_c\ll 1$ in the same formulation, but it is not meaningful to use $\mu$ for super-critical conditions because $\psi_s$ does not exist. Thus we introduce $\nu\ll 1$ such that $\delta=\delta_c(1\pm\nu)$, with plus and minus signs, respectively, corresponding to super- and sub-critical regimes. A correspondence between $\nu\ll 1$ and $\mu\ll 1$ can be made from~\eqref{steady1} for the sub-critical case, $2\nu= \mu^2 + O(\mu^{3})$.

The analysis of the near-critical limit, originally due to Frank-Kamenetskii~\cite{frank1946}, has been employed recently for premixed reactants flowing inside a pipe~\cite{moreno2017}.  The required rescalings for the current problem are found to be $\xi=\pi\nu\tau L^2/(2R^2)$ and $\psi = 1 + \sqrt{2\nu} \chi - \nu$ (cf.~\cite{frank1969}, p.~406 and~\cite{zeldovich1985}, p.~68). Substituting these new variables into~\eqref{intfinal} and neglecting terms of order $\nu^2$ leads to
\begin{equation}
    \int_0^\xi \frac{\dd \chi}{\dd\xi'} \frac{\dd\xi'}{\sqrt{\xi-\xi'}} =  \chi^2 \pm 1, \qquad \chi(0)\rightarrow -\frac{1-\psi_o}{\sqrt{2\nu}}\label{nearcrit}
\end{equation}
where $\psi_o<1$ is an order-unity constant that contains information from the initial periods. In the sub-critical case,  solution for $\chi$ exists at all times approaching $\chi+1\sim 1/\sqrt{\xi}$ as $\xi\rightarrow\infty$. On the contrary, for $\delta>\delta_c$, $\chi$ diverges at a finite value $\xi_c$ of $\xi$ according to $\chi\sim 1/\sqrt{\xi_c-\xi}$. The value of $\xi_c$  depends on $\nu$, and preliminary numerical integrations suggested that $\xi_c\sim (2\nu)^{-1}$.

 Had we considered the case $\gamma\neq 0$, an additional term $\gamma (2\nu\xi)^{-1/2}$ would appear on the left-hand side of equation~\eqref{nearcrit}, suggesting two possible modifications for the super-critical case. For $\gamma>0$, a balance of this new term with the Arrhenius term leads to $\chi(\xi)\rightarrow-[\gamma^2/(2\nu\xi)]^{1/4}\gg \chi(0)$ (only the negative root is allowed since only then will the quadratic approximation in~\eqref{nearcrit} be a representative of the full Arrhenius expression) indicating that the solution has not yet approached the steady solution at this time scale. On the other hand, for $\gamma<0$, we cannot find any real solution, meaning that the blowup of $\chi$ has already taken place at a time scale smaller than $\xi$. Different scalings therefore would be needed for $\gamma^2\gtrapprox 2\nu$.

\section{General discussions}\label{sec:discussion}

 From~\eqref{critical}, in view of~\eqref{nondim1} and~\eqref{nondim2}, explicit expressions are available for the critical Damk\"ohler numbers above which ignition occurs, written in dimensional variables as
 \begin{equation}
     \frac{aq_s E}{\bar R T_{Io}'^2}e^{-E/\bar RT_{Io}'} \geq \frac{\lambda_O}{e} \exp\left[\frac{Er_F(T_{Fo}'-T_{Oo}')}{\bar R T_{Io}'^2 (r_F+r_O)}\right], \label{criticaldim}
 \end{equation}
 where $T_{Io}'=(r_FT_{Fo}'+ r_O T_{Oo}')/(r_F+r_O)$. For fixed initial temperatures, the phenomenon of ignition is favored by a larger fuel radius $a$ or a larger reaction parameter $q_se^{-E/\bar R T_{Io}'}E/(\bar R T_{Io}')$. Decreasing $a$ delays ignition and makes it less likely to occur because it enhances (geometrically) rates of heat conduction away from the interface where the heat is being liberated. \blue{This is in qualitative agreement with the droplet-droplet collision experiments~\cite{williamsreport,zhang2016}, where it was observed that the ignition delay time increases with increasing droplet size ratio, defined such that this ratio is greater than unity}. The augmented heat-conduction rates for small fuel-droplets undermine the effects of thermal conductivity of the fuel $\lambda_F$ (as if $\lambda_F$ is infinite), leaving heat-loss to the oxidizer as the principal way of taking heat away from the interface; this explains the presence of $\lambda_O$, but not $\lambda_F$ in the critical condition. Next keeping fuel radius and reaction parameter fixed, ignition is favored more for the case $T_{Fo}'<T_{Oo}'$ since now there is an adverse temperature gradient (additional to the gradient-jump induced by the interface reaction) at the interface that opposes heat-conduction to the oxidizer. The opposite case $T_{Fo}'>T_{Oo}'$ increases heat-loss to the oxidizer making it harder for the interface temperature to blow-up. The thermal responsivity is closely associated with the fact that the initial temperatures of the two reactants are different from each other and it disappears in the critical condition (or, in the ignition formula given below) whenever $T_{Fo}'=T_{Oo}'=T_{Io}'$. Interestingly enough, when $T_{Fo}'\neq T_{Oo}'$, only the ratio $r_O/r_F$ appears explicitly in the critical condition as well as affects the initial interface temperature. If the effect on $T_{Io}'$ is ignored, then increasing this ratio promotes the ignition whenever $T_{Fo}'>T_{Oo}'$.  
  
The ignition delay time in dimensional form that is appropriate for large Damk\"ohler number \blue{(or, equivalently when the difference in the inequality~\eqref{criticaldim} is large)} defined in section VI is
\begin{equation}
    t_{\rm{ign}} = \frac{c}{\pi} \left[\frac{(r_F+r_O)\bar R T_{Io}'^2}{q_s E e^{-E/\bar RT_{Io}'}}\right]^2, \label{tign1}
\end{equation}
where the value of $c(\Lambda,k)$ can be obtained from Figure~\ref{fig:cetap} for given values of
\begin{equation}
    \Lambda = \frac{(T_{Fo}'-T_{Oo}')(r_F\lambda_O + r_O \lambda_F)}{aq_s(r_F+r_O) e^{-E/\bar RT_{Io}'}} \quad \text{and} \quad k = \frac{(\lambda_O-\lambda_F)\bar R T_{Io}'^2}{a q_sE e^{-E/\bar RT_{Io}'}}.
\end{equation}
This ignition formula is same as that for the planar problem. As expected, the dependence of the ignition delay time on the fuel radius $a$ entering implicitly via $\Lambda$ and $k$, disappears in the planar limit $a\rightarrow \infty$. In the planar limit, the ignition formula is independent of the thermal conductivities because the heat-conduction is driven by huge temperature gradients induced by the fast chemical reaction, on both sides of the interface and not by the values of the thermal conductivities; the difference $\lambda_O-\lambda_F$ emerges in the first-order correction to the planar problem. Similarly, the temperature gradients associated with the initial-temperature differences also arise only in the first-order correction. The ignition formula, however, does depend on the sum of the responsivities since although the interface temperature is continuous at the interface, the temperature on either sides suffer rapid variations (quite independent of all parameters except the reaction parameter) that necessitate the symmetric presence of the thermal responsivities in the formula.

In the near-critical limit, the ignition delay time can be written as
\begin{equation}
    t_{\rm{ign}} = \frac{b}{\pi}\left[\frac{a r_O \bar R T_{Io}'^2 }{aeq_sE e^{-E/\bar RT_{Io}'} - \lambda_O \bar R T_{Io}'^2}\right]^2,
\end{equation}
where $b$ is an order-unity constant that is not determined in this paper. Since the preceding formula reduces to ~\eqref{tign1} as $a\rightarrow \infty$, the formula can be anticipated to be of general validity, although $b$ may no longer be a universal constant except for the two extreme cases considered in this paper. The ignition delay time is inversely proportional to the measure of the inequality in equation~\eqref{criticaldim} (after setting $T_{Fo}'=T_{Oo}'$) and consequently discussions pertaining to~\eqref{criticaldim} are also valid here. For example, while the leading-order large Damk\"ohler-number ignition formula~\eqref{tign1} is symmetric with respect to both fuel and oxidizer, the above near-critical ignition formula depends only on the thermal properties of the oxidizer and not on that of the fuel.

\section{Conclusions}\label{sec:conclusion}

The principal result of the paper is the determination of critical conditions that separate steady solutions from thermal runaway. It is observed that ignition is promoted by increasing the fuel radius, or the reaction parameter or the initial temperature of the oxidizer or by decreasing the oxidizer thermal conductivity. \blue{Whenever the initial temperatures of the two reactants are not the same, the thermal responsivity ratio, in part, determines whether the chance of igniting the reactants is enhanced or worsened.} In addition, the ignition delay times for large and near-critical Damk\"ohler numbers are also estimated, where in the former case, the delay time is symmetric about the two reactants, while for the near-critical conditions, the delay time is dependent only on the properties of the oxidizer. In general, the ignition delay time is inversely proportional to the reaction parameter and directly proportional to the thermal responsivities, starting from the case where the dependence is only on the oxidizer thermal responsivity to the case where the dependence is on both the fuel and the oxidizer thermal responsivities in a symmetric fashion. Knowledge of these dependences on various parameters are critical in having a better control on hypergolic ignition.

However the paper mainly addresses the influences of finite proportions of the reactants and clearly more work is required for further understanding of hypergolic ignition and to make concrete predictions; in particular, to name a few of the factors that are neglected here, there may be interest in the effects of reactant depletion, the effects of detailed chemical kinetics, the effects of gel additives etc. An ignition study employing a reduced chemistry for the hypergolic reactants, but more complex than the one-step chemistry considered here could be the next step for future investigations.

\section*{Acknowledgments}
The author is indebted to Professor Forman A. Williams for the helpful remarks and various suggestions on the manuscript; in particular, for his encouragement in publishing this work. The author would like to thank Professor Antonio L. S\'anchez for the continuous support and for pointing out the near-critical limit. Finally, the author is also most grateful to Adam D. Weiss for general discussions on various aspects of the problem.

\bibliography{sample}

\end{document}